\documentstyle[12pt,aps,bezier]{revtex}
\begin{document}
\draft
\title{\bf Magnetic Oscillations in the Nambu -- Jona-Lasinio
model}
\author{{\bf D. Ebert$^\ast$ and K.G. Klimenko$^\dagger$}}
\address{{\it $^\ast$Institut f\"ur Physik,
Humboldt-Universit\"at,
D-10115 Berlin, Germany}}
\address{{\it $^\dagger$Institute for High Energy Physics,
142284
Protvino, Moscow Region, Russia}}
\vspace{0.5cm}

\maketitle
\begin{abstract} {\bf Abstract.}
The phase structure of a simple Nambu--Jona-Lasinio model has
been investigated at non-zero values of $\mu$ and $H$, where $H$
is an external magnetic field and $\mu$ is the chemical
potential.  On this basis magnetic oscillations effects were
considered. It was shown that there are standard (periodic) van
Alphen--de Haas magnetic oscillations of some thermodynamical
quantities, including magnetization, pressure and particle
density in the NJL system.  Besides, we have found non-standard,
i.e.  non-periodic, magnetic oscillations, since the frequency
of oscillations is a $H$-dependent quantity.  Finally, there
arises an oscillating behaviour not only for thermodynamical
quantities, but also for a dynamical quantity like the quark
mass.
\end{abstract}
\vspace{1cm}

Magnetic oscillations effects are well-known phenomena in
condensed matter physics. In particular, the oscillation effect
of the magnetization, which is called now  the van Alphen--de
Haas effect,  was for the first time predicted by Landau and
then experimentally observed in some non-relativistic systems
(in metals) more than sixty years ago \cite{lan,haas}.  At
present, a lot of the attention of researchers dealing with
magnetic oscillations is focused on relativistic condensed
matter systems (mainly on QED at nonzero values of the chemical
potential $\mu$ and external magnetic field $H$), since the
results of these studies may be applied to cosmology,
astrophysics and high energy physics
\cite{per,as}.

 It was shown in the framework of QED that the thermodynamical
potential $\Omega(\mu,H)$ of the system has in 1-loop
approximation the following form 
$
\Omega(\mu,H)=\Omega_{mon}(\mu,H)+\Omega_{osc}(\mu,H),
$
where $\Omega_{mon}(\mu,H)$ is the  monotonic part of
$\Omega(\mu,H)$, and all magnetic oscillations are contained in
the so-called oscillating part
\begin{equation}
\label{eq.100}
\Omega_{osc}(\mu,H)=\sum_{k=1}^\infty [A_k(H)\cos(2\pi k\omega)+
B_k(H)\sin(2\pi k\omega)],
\end{equation}
where $\omega=(\mu^2-m^2)/(2eH)$ ($e,m$ are electric charge and
mass of fermions, respectively), and $A_k(H),B_k(H)$ are
smoothly varying functions.  Due to the presence of
trigonometric functions, expression (\ref{eq.100}) obviously
oscillates over the variable $(2eH)^{-1}$ with the frequency
$(\mu^2-m^2)$, which is not an $H$-dependent quantity. In
condensed matter physics such kind of oscillations are usually
called  periodic ones.

In the present talk magnetic oscillation effects are considered
in
the framework of quantum field theory with four-fermion 
interactions 
\begin{equation}
\label{eq.1}
L=\sum_{k=1}^N\bar{q}_k i \hat {\partial} q_k +\frac{G}{2N}
[(\sum_{k=1}^N\bar{q}_k
q_k)^2 +(\sum_{k=1}^N\bar{q}_k i\gamma_5 q_k)^2],
\end{equation}
which is the $N$-fermionic extension of the simplest Nambu --
Jona-Lasinio model (NJL) \cite{nam}.~\footnote{For simplicity,
we consider in the following fermions (``quarks'') of equal
electric charge.} Obviously, the model (\ref{eq.1}) is invariant
under (global) $SU(N)$ and $U(1)_V$ transformations as well as
continuous $U(1)_A$ chiral transformations: $q_k\to
e^{i\theta\gamma_5}q_k~~$;$~~ (k=1,...,N).$

We shall find the thermodynamic potential $\Omega(\mu,H)$, which
is related to the corresponding effective potential $V_{H\mu}
(\Sigma)$  of the NJL system (\ref{eq.1}) by
\begin{equation}
\label{eq.101}
\Omega(\mu,H)=V_{\mu H}(\Sigma)~
\vrule_{~~ \atop \scriptstyle{\Sigma=\Sigma_{min}}}
\end{equation}
and contains all the information about thermodynamical
quantities such as magnetization, particle density, etc.  In the
relation (\ref{eq.101}), one should first of all calculate the
effective potential $V_{H\mu}(\Sigma)$. So, before considering
the magnetic oscillations, we can study the vacuum properties of
the NJL model.

Notice that special attention has been paid to the analysis of
the vacuum structure of NJL-type models at non-zero temperature
and chemical potential
\cite{kaw,k1}, in the presence of external (chromo-)magnetic
fields \cite{klev,eb,gus}, with allowance for curvature and
non-trivial space-time topology \cite{In,k2}. The combined
influence of external electromagnetic and gravitational fields
on the dynamical chiral symmetry breaking (DCSB) effect in
four-fermion field theories was investigated in \cite{muta,od}.
However, the influence of both an external magnetic field $H$
and chemical potential $\mu$ on the phase structure of the NJL
model was not considered up to now.

\underline{\bf Phase structure of the model.} The necessary
information about the phase structure of a given field
theoretical model is contained in the global minimum point of
the corresponding effective potential.  In the presence of
$\mu,H$ the effective potential $V_{H\mu}(\Sigma)$ of the NJL
model has in  leading order of large $N$ the following form
\begin{equation}
\label{eq.3}
V_{H\mu}(\Sigma)=V_H(\Sigma)-\frac{eH}{4\pi^2}\sum_{k=0}^{\infty
}
\alpha_k
\theta (\mu-s_k)\Biggl\{\mu\sqrt{\mu^2-s_k^2}-s_k^2\ln\left
[\frac {\mu+\sqrt{\mu^2-s_k^2}}{s_k}\right ]\Biggr\},
\end{equation}
where $\alpha_k=2-\delta_{k0}$, $s_k=\sqrt{\Sigma^2+2eHk}$.
$V_H(\Sigma)$ is the effective potential at $\mu =0$, $H\not =0$
\begin{equation}
\label{eq.4}
V_H(\Sigma)=\frac{H^2}2+V_0(\Sigma)-\frac{(eH)^2}{2\pi^2}\Bigl\{
\zeta
'(-1,x)-\frac 12[x^2-x]\ln x +\frac{x^2}4\Bigr\},
\end{equation}
where $x=\Sigma^2/(2eH)$, $\zeta (\nu,x)$ is the generalized
Riemann zeta-function, $\zeta'(-1,x)$$~=
d\zeta(\nu,x)/d\nu|_{\nu=-1}$, and
\begin{equation}
V_0(\Sigma)=
\label{eq.5}
\frac{\Sigma^2}{2G}-
\frac{1}{16\pi^2}\Biggl\{
\Lambda^4\ln
\left(1+\frac{\Sigma^2}{\Lambda^2}\right)+\Lambda^2\Sigma^2
-\Sigma^4\ln
\left(1+\frac{\Lambda^2}{\Sigma^2}\right)\Biggr\}
\end{equation}
is the effective potential at $H,\mu =0$. In (\ref{eq.5})
$\Lambda$ is the ultraviolet cut off parameter. Finally, let us
remark that $\Sigma$ is an auxiliary scalar field, which, at the
tree level, is proportional to $\bar qq$ by the equations of
motion. The global minimum point of the potential (\ref{eq.3})
defines the vacuum expectation value of $\Sigma$ and is equal to
the dynamical quark mass.

At $\mu,H=0$ and $G<G_c=4\pi^2/\Lambda^2$ the global minimum
point of $V_0(\Sigma)$ equals to the value $\Sigma=0$. Hence, in
this case quarks are massless and chiral symmetry remains
intact. If $G>G_c$, the effective potential (\ref{eq.5}) has a
nontrivial global minimum point, which we shall denote as $M$.
(Evidently,  $M$ depends on the values of $G$ and $\Lambda$
\cite{k1}.)

At $\mu=0,~ H\not =0$ the chiral symmetry of the model is
spontaneously broken for arbitrary values of the bare coupling
constant $G$. This is due to the fact, that the global minimum
point $\Sigma_0(H)$ of the potential $V_H(\Sigma)$ is unequal to
zero \cite{klev,gus}.

 In order to study the properties of the NJL model vacuum for
the general case, when both $\mu$ and $H$ are nonzero, one
should find all solutions of the stationarity equation
\begin{equation}
\label{eq.6}
\frac {\partial}{\partial\Sigma}V_{H\mu}(\Sigma)=\frac
 {\partial}{\partial\Sigma}V_{H}(\Sigma)+
\frac{2eH\Sigma}{4\pi^2}\sum_{k=0}^{\infty} \alpha_k \theta
(\mu-s_k)\ln\left
[\frac {\mu+\sqrt{\mu^2-s_k^2}}{s_k}\right ]=0
\end{equation}
and select that one, at which the potential $V_{H\mu}(\Sigma)$
takes its smallest value. This is the global minimum point for
the function (\ref{eq.3}). The properties of this point as a
function of $\mu$ and $H$ give us  a lot of information about
the ground state.  We omit here the detailed consideration of
this procedure and present directly the phase structure
description of the model  (Figure 1).

\begin{figure}[t]
\begin{center}
\unitlength=1.0mm
\special{em:linewidth 0.4pt}
\linethickness{0.4pt}
\begin{picture}(147.00,132.67)
\put(4.67,15.00){\vector(1,0){136.67}}
\put(14.33,7.67){\vector(0,1){125.00}}
\put(14.33,23.67){\circle*{1.33}}
\put(14.33,23.67){\line(1,0){6.67}}
\put(23.33,23.67){\line(6,1){8.00}}
\put(33.33,24.67){\line(6,1){8.33}}
\put(44.33,26.67){\line(6,1){8.00}}
\put(55.00,28.67){\line(4,1){9.00}}
\put(66.67,31.33){\line(4,1){8.33}}
\put(77.33,33.67){\line(4,1){8.33}}
\put(88.00,36.00){\line(2,1){7.33}}
\put(97.33,40.33){\line(5,3){7.33}}
\put(106.67,45.67){\line(2,1){6.67}}
\put(116.00,50.00){\line(4,1){8.33}}
\put(126.67,53.33){\line(-1,-2){0.33}}
\put(126.33,52.67){\line(6,1){7.00}}
\put(136.00,54.33){\line(6,1){7.33}}
\put(107.00,45.67){\circle*{1.00}}
\put(93.67,67.33){\circle*{1.00}}
\put(81.67,69.67){\circle*{1.00}}
\put(68.33,67.67){\circle*{1.00}}
\put(57.67,70.67){\circle*{1.00}}
\put(46.33,69.67){\circle*{1.00}}
\put(37.33,69.33){\circle*{1.00}}
\put(30.33,70.67){\circle*{1.00}}
\put(23.33,72.33){\circle*{0.00}}
\put(19.33,71.67){\circle*{0.00}}
\put(16.33,70.67){\circle*{0.00}}
\put(14.33,70.33){\circle*{1.33}}
\put(93.67,67.33){\line(-1,2){2.00}}
\put(81.67,69.67){\line(-4,-5){4.67}}
\put(75.67,62.00){\line(-6,-5){6.00}}
\put(68.33,55.33){\line(-6,-5){6.33}}
\put(60.00,48.00){\line(-3,-2){6.67}}
\put(51.00,42.00){\line(-5,-3){7.67}}
\put(40.67,35.33){\line(-2,-1){8.00}}
\put(30.67,29.67){\line(-5,-2){6.67}}
\put(21.33,25.67){\line(-4,-1){7.00}}
\put(68.33,67.67){\line(-2,5){1.67}}
\put(57.67,70.33){\line(-2,-3){3.33}}
\put(53.33,63.33){\line(-3,-5){3.67}}
\put(48.00,55.67){\line(-3,-4){4.33}}
\put(42.00,47.67){\line(-4,-5){4.67}}
\put(35.67,40.00){\line(-1,-1){4.67}}
\put(29.00,33.67){\line(-5,-4){6.33}}
\put(21.00,27.67){\line(-5,-3){6.67}}
\put(37.33,69.33){\line(-2,-5){3.00}}
\put(33.67,60.33){\line(-1,-2){3.67}}
\put(28.67,50.33){\line(-2,-5){2.67}}
\put(24.67,41.33){\line(-1,-2){4.00}}
\put(19.67,31.00){\line(-3,-4){5.67}}
\put(93.67,67.33){\line(5,3){41.33}}
\put(68.33,67.33){\line(1,1){44.33}}
\put(46.33,69.33){\line(3,5){27.33}}
\put(30.33,70.67){\line(1,4){11.33}}
\put(137.00,10.00){\makebox(0,0)[cc]{$\sqrt{eH}$}}
\put(8.33,127.00){\makebox(0,0)[cc]{$\mu$}}
\put(8.33,69.67){\makebox(0,0)[cc]{$M_1$}}
\put(9.33,23.67){\makebox(0,0)[cc]{$M$}}
\put(95.00,25.33){\makebox(0,0)[cc]{$B$}}
\put(73.67,42.67){\makebox(0,0)[cc]{$C_0$}}
\put(51.33,49.00){\makebox(0,0)[cc]{$C_1$}}
\put(36.00,50.00){\makebox(0,0)[cc]{$C_2$}}
\put(23.33,51.00){\makebox(0,0)[cc]{$C_3$}}
\put(19.67,51.33){\circle*{0.00}}
\put(17.00,51.00){\circle*{0.00}}
\put(124.33,72.00){\makebox(0,0)[cc]{$A_0$}}
\put(101.67,84.67){\makebox(0,0)[cc]{$A_1$}}
\put(75.33,93.33){\makebox(0,0)[cc]{$A_2$}}
\put(48.00,95.33){\makebox(0,0)[cc]{$A_3$}}
\put(28.33,95.00){\circle*{0.00}}
\put(23.33,95.00){\circle*{0.00}}
\put(18.00,95.00){\circle*{0.00}}
\put(134.67,49.67){\makebox(0,0)[cc]{$\mu_c$}}
\put(133.33,94.33){\makebox(0,0)[cc]{$l_1$}}
\put(109.67,114.33){\makebox(0,0)[cc]{$l_2$}}
\put(74.67,117.67){\makebox(0,0)[cc]{$l_3$}}
\put(42.67,120.33){\makebox(0,0)[cc]{$l_4$}}
\put(109.33,43.33){\makebox(0,0)[cc]{$t_0$}}
\put(90.33,64.67){\makebox(0,0)[cc]{$s_0$}}
\put(81.00,73.00){\makebox(0,0)[cc]{$t_1$}}
\put(65.67,65.33){\makebox(0,0)[cc]{$s_1$}}
\put(57.33,73.67){\makebox(0,0)[cc]{$t_2$}}
\put(44.33,66.67){\makebox(0,0)[cc]{$s_2$}}
\put(37.00,73.00){\makebox(0,0)[cc]{$t_3$}}
\put(28.00,67.67){\makebox(0,0)[cc]{$s_3$}}
\put(57.67,70.67){\line(5,6){3.00}}
\put(30.33,70.67){\line(-5,2){4.33}}
\linethickness{1.0pt}
\bezier{116}(107.00,45.67)(96.00,51.33)(93.67,67.33)
\bezier{84}(81.67,69.33)(69.00,61.33)(68.33,67.33)
\bezier{72}(57.67,70.33)(47.67,63.67)(46.33,69.33)
\bezier{52}(37.67,69.00)(31.00,65.00)(30.33,70.33)
\bezier{12}(37.33,69.33)(38.67,71.00)(40.00,72.00)
\bezier{12}(46.33,69.67)(45.67,71.33)(45.00,72.67)
\bezier{12}(41.00,73.00)(42.00,74.33)(43.67,73.67)
\bezier{28}(61.67,74.67)(63.67,76.67)(66.33,72.67)
\bezier{16}(81.67,69.33)(83.33,71.33)(84.33,72.33)
\bezier{28}(85.67,73.33)(88.67,75.00)(90.67,72.67)
\end{picture}\\
{\footnotesize {\bf FIGURE 1}. Phase structure of the NJL model.
(Detailed description of the figure is given in the text.)}
\end{center}
\end{figure}

In this figure, in the plane $(\mu,\sqrt{eH})$ the phase
portrait of the model is qualitatively represented for the case
$G_c<G<(1.225...)G_c$, where $M$ is the quark mass at $\mu
=H=0$,
$M_1=(\Lambda^2/2-2\pi^2/G)^{1/2}$. Here one can
see infinite sets of symmetric massless $A_0,A_1,...$ phases, as
well as massive phases $C_0,C_1,...$ with DCSB. In addition,
there is another massive phase $B$.  Dashed and solid lines in
Figure 1 are critical curves of first- and second-order phase
transitions, respectively.  One can also see on this phase
portrait infinitely many tricritical points $t_k,s_k$
($k=0,1,2,...$) which lie on the boundary between massless and
massive phases (chiral boundary). (A point of the phase diagram
is called a tricritical one if, in an arbitrarily small vicinity
of it, there are first- as well as second-order phase
transitions.) Numerical investigation gives the following values
of the external magnetic field corresponding to tricritical
points $t_0$ and $s_0$ at different values of the bare coupling
constant $G$: $eH_{t_0}/\Lambda^2$=$0.01...;~0.08...;~0.13...$
as
well as $eH_{s_0}/\Lambda^2$=$0.006...;~0.056...;~0.103...$
 for $G/G_c$=$1.01;~1.1;~1.2$, respectively.  We should also
remark that the part $\overbrace{t_0\mu_c(H)}$ of the chiral
boundary is described by the equation
$V_{H\mu}(0)=V_{H\mu}(\Sigma_0(H))$.

Points $(\mu,H)$ of the phase diagram, lying above the chiral
boundary, correspond to the chirally symmetric ground state of
the NJL model. One-fermion excitations of this vacuum have zero
masses. At first sight, it might seem that the properties of
this symmetric vacuum are slightly varied, when parameters $\mu$
and $H$ are changed. However, this is not the case, and in this
region, as was mentioned above, we have infinitely many massless
symmetric phases of the theory corresponding to infinitely many
Landau levels, as well as a variety of critical curves of
second-order phase transitions. Let us next show this.

It is well-known that the state of thermodynamic equilibrium
(the ground state) of an arbitrary quantum system is described
by the thermodynamic potential (TDP) $\Omega$, which is just the
value of the effective potential at its global minimum point
(see (\ref{eq.101})).  In the case under consideration, the TDP
$\Omega(\mu,H)$ at $\mu>M_1$ (see  Figure 1) has the form
\begin{eqnarray}
\Omega(\mu,H)&\equiv&V_{H\mu}(0)=V_H(0)-  \nonumber \\
&&- \frac{eH}{4\pi^2}\sum^{\infty}_{k=0}\alpha_k\theta
(\mu-\epsilon_k)\Bigl\{\mu\sqrt{\mu^2-\epsilon_k^2}-\epsilon_k^2
\ln
\left[\left(\sqrt{\mu^2-\epsilon_k^2}
+\mu\right)/\epsilon_k\right]\Bigr\},
\label{eq.25}
\end{eqnarray}
where $\epsilon_k=\sqrt{2eHk}$. We shall use the following
criterion of phase transitions: if at least one first (second)
partial derivative of $\Omega(\mu,H)$ is a discontinuous
function at some point, then this is a point of a first-
(second-) order phase transition.

Using this criterion, let us show that lines
$l_k=\{(\mu,H):\mu=\sqrt{2eHk}\}$ ($k=1,2,...$), are critical
lines of second-order phase transitions.

Indeed, from (\ref{eq.25}) one easily finds
\begin{equation}
\label{eq.27}
\frac {\partial\Omega}{\partial\mu}\bigg |_{(\mu,H)\to l_{k+}}-
\frac {\partial\Omega}{\partial\mu}\bigg |_{(\mu,H)\to
l_{k-}}=0,
\end{equation}
as well as:
\begin{equation}
\label{eq.28}
\frac {\partial^2\Omega}{(\partial\mu)^2}\bigg |_{(\mu,H)\to
l_{k+}}-
\frac {\partial^2\Omega}{(\partial\mu)^2}\bigg
|_{(\mu,H)\to l_{k-}}= -\frac
{eH\mu}{2\pi^2\sqrt{\mu^2-\epsilon_k^2}}\bigg |_{\mu\to
\epsilon_{k+}}\rightarrow -\infty.
\end{equation}
Equation (\ref{eq.27}) means that the first derivative
$\partial\Omega/\partial\mu$ is a continuous function on all
lines $l_k$. However, the second derivative
$\partial^2\Omega/(\partial\mu)^2$ has an infinite jump on each
line $l_k$ (see (\ref{eq.28})), so these lines are critical
curves of second-order phase transitions.  (Similarly, we can
prove the discontinuity of $\partial^2\Omega/(\partial H)^2$ and
$\partial^2\Omega/\partial\mu\partial H$ on all lines $l_n$.)

The presence of an infinite set of massive phases $C_k$ on the
phase portrait is conditioned by a special structure of
the stationarity equation (\ref{eq.6}). Analytical and numerical
considerations of it show that below the chiral boundary the
effective potential global minimum point $\Sigma(\mu,H)$, which
is identical to the quark mass, has $\mu$ and $H$ dependences.
The function $\Sigma(\mu,H)$ is a continuous one inside each of
regions $C_k$. However, it is a discontinuous one on each of the
curves $\overbrace{Mt_{k}}$, where the quark mass changes its
value by a jump.  That is why boundaries between $C_k$-regions
are the first order phase transition lines. In contrast, in the
phase $B$, the global minimum point is equal to $\Sigma_0(H)$
($\equiv$ quark mass in the case $\mu=0,H\not =0$), which is a
$\mu$-independent quantity. This means that the particle density
$n\equiv -\partial\Omega/\partial\mu$ in the ground state of
the phase $B$ is identically equal to zero, whereas in each
phase
$C_k$ this quantity differs from zero.

\underline{\bf Magnetic oscillations.}
Now we want to show that there arise, from the presence of
infinite sets of massless $A_k$ phases as well as of massive
$C_k$ ones, magnetic oscillations (the so-called van Alphen--de
Haas-type effect) of some physical parameters in the NJL model
gauged by an external magnetic field.

Let the chemical potential be fixed, i.e.  $\mu = \rm {const} >
M_1$ (see Figure 1).  Then on the plane $(\mu,\sqrt{eH})$ (see
Figure 1) we have a line that crosses the critical lines
$l_1,l_2,...$ at points corresponding to some values
$H_1,H_2,...$ 
of the external magnetic field. The particle density $n$
and the magnetization $m$ of any thermodynamic system are
defined by the TDP in the following way:
$n=-\partial\Omega/\partial\mu$, $m=-\partial\Omega/\partial H$.
At $\mu$ = const these quantities are continuous functions of
the external magnetic field only, i.e. $n\equiv n(H),~m\equiv
m(H)$. We know that all the second derivatives of
$\Omega(\mu,H)$ are discontinuous on every critical line $l_n$
(see (\ref{eq.28})). The functions $n(H)$ and $m(H)$, being
continuous in the interval $H\in (0,\infty)$, therefore have
first derivatives that are discontinuous on an infinite set of
points $H_1,...,H_k,...$ Such a behaviour manifests itself a
phenomenon usually called oscillations.

Analogously to QED and condensed-matter physics \cite{lan,haas}, 
let us again separate the expression for a physical quantity
with
oscillations into two parts: the first  monotonic one
 does not contain any oscillations, whereas the second part,
which is of particular interest here, contains all the
oscillations. Following this rule, we can write down, say, the
TDP (\ref{eq.25}) of the NJL model in the form
$
\Omega(\mu,H)=\Omega_{mon}(\mu,H)+\Omega_{osc}(\mu,H).
$
In order to present the oscillating part $\Omega_{osc}(\mu,H)$
in an
analytical form, we shall use the technique elaborated in
\cite{as}, where manifestly analytical expressions for this
quantity were found in the case of a perfect relativistic
electron--positron gas.  This technique can be used without any
difficulties in our case, too.  So, applying in (\ref{eq.25})
the Poisson summation formula \cite{lan}
\begin{equation}
\label{eq.30}
\sum^{\infty}_{n=0}\alpha_n\Phi (n)
=2\sum^{\infty}_{k=0}\alpha_k
\int\limits^{\infty}_{0}\Phi (x)
\cos (2\pi kx)dx,
\end{equation}
where $\alpha_n=2-\delta_{n0}$, one can get for
$\Omega_{osc}(\mu,H)$
the following expression
\begin{equation}
\label{eq.32}
\Omega_{osc}=\frac{\mu}{4\pi^{3/2}}
\sum^{\infty}_{k=1}
\left (\frac{eH}{\pi k}\right)^{3/2}
[Q(\pi k \nu)\cos (\pi k\nu+\pi/4)+ P(\pi k \nu)\cos (\pi k\nu
-\pi/4)],
\end{equation}
where $\nu=\mu^2/(eH)$. The functions $P(x)$ and $Q(x)$
in (\ref{eq.32}) are connected with the Fresnel
integrals $C(x)$ and $S(x)$ \cite{k15}:~
$C(x)=\frac{1}{2}+\sqrt{\frac{x}{2\pi}} [P(x)\sin x +Q(x)\cos
x]$,~$S(x)=\frac{1}{2}-\sqrt{\frac{x}{2\pi}} [P(x)\cos x -Q
(x)\sin x].$
They have, at $x\to\infty$, the following asymptotics
\cite{k15}:
$
P(x)=x^{-1}-3x^{-3}/4+...$,$~Q(x)=-x^{-2}/2+
15x^{-4}/8+...$
Formula (\ref{eq.32}) presents, in a manifestly analytical form,
the oscillating part of the TDP (\ref{eq.25}) for the NJL model
at $\mu>M_1$. In the case under consideration, since the TDP is
proportional to the pressure of the system, one can conclude
that the pressure in the NJL model oscillates when $H\to 0$,
too.  It follows from (\ref{eq.32}) that the frequency of
oscillations over the parameter $(eH)^{-1}$ equals $\mu^2/2$
and does not depend on $H$. So, in this case we have periodic
magnetic oscillations.
Then, starting from (\ref{eq.32}), one can easily find the
corresponding expressions for the oscillating parts of $n(H)$
and $m(H)$. These quantities oscillate at $H\to 0$ with the same
frequency $\mu^2/2$ and have a rather involved form, so we do
not present them here.

Finally, we should note that the character of magnetic
oscillations in the NJL model at $\mu>M_1$ resembles the
magnetic
oscillations in massless quantum electrodynamics \cite{per,as}.
Indeed, in this case in both models one can find periodic
magnetic
oscillations of some thermodynamic parameters.

Now let us show that at a fixed value of the chemical
potential and $M<\mu<M_1$ the character of magnetic
oscillations is changed. In this case on the plane
$(\mu,\sqrt{eH})$
we 
have a line drawn through an infinite set of the $C_k$-phases.
Hence, the thermodynamic potential of the NJL system 
has the following form:
$\Omega(\mu,H)=$$V_{H\mu}(\Sigma(\mu,H))$,
where $\Sigma(\mu,H)$ is the global minimum point of the 
potential $V_{H\mu}(\Sigma)$. Applying in (4) again
the formula (\ref{eq.30}), one can find  the following
expression
for the oscillating part of TDP
\begin{equation}
\label{eq.11}
\Omega_{osc}\sim
\sum^{\infty}_{k=1}
\left (\frac{eH}{\pi k}\right)^{3/2}
[Q(\pi k\nu)\cos (2\pi k\omega+\pi/4)
+P(\pi k\nu)\cos (2\pi k\omega -\pi/4)],
\end{equation}
where $\nu=\mu^2/(eH)$, $\omega=(\mu^2-\Sigma^2(\mu,H))/(2eH)$.
From (\ref{eq.11}) one can see that the TDP $\Omega(\mu,H)$
oscillates
with
frequency $(\mu^2-\Sigma^2(\mu,H))/2$ if the variable
$(eH)^{-1}$
tends to
infinity.  Since $\Omega(\mu,H)$ is, up to a sign, equal to the
pressure in the ground state of the system, also in the present
case the pressure in the NJL model is an oscillating quantity.
Moreover, other thermodynamic quantities such as
particle density $n=-\partial\Omega/\partial\mu$ and
magnetization $m=-\partial\Omega/\partial H$ oscillate with the
same
frequency.

Here we should do an  important remark.
In the NJL model at $M<\mu<M_1$, in contrast to QED,
the magnetic oscillation frequency is a $H$-dependent quantity.
(Since the quark mass $\Sigma(\mu,H)$ has $H$-dependency.) So,
strictly speaking, in the NJL model magnetic oscillations are
not periodic ones.  Recently, similar peculiarities of magnetic
oscillations are observed in some ferromagnetic semiconductive
materials such as $\rm HgCr_2Se_4$
\cite{bal}, where non-periodic magnetic oscillations over the
variable
$(eH)^{-1}$ were found to exist for electric
conductivity as well as magnetization.

Finally, we should remark that in the NJL model not only
thermodynamic quantities oscillate, but some dynamical
parameters of the system do as well. This concerns, in
particular, oscillations of the dynamical quark mass. In fact,
by applying the Poisson summation formula (\ref{eq.30}) to the
stationarity equation (\ref{eq.6}) and searching for the
solution $\Sigma(\mu,H)$ of this equation in the form
$\Sigma(\mu,H)=\Sigma_{mon}+\Sigma_{osc}$, one can easily find
the following expressions for $H\to 0$:
\begin{equation}
\Sigma_{osc}(\mu,H)\sim
\frac {(eH)^{3/2}}{\mu\tilde M}
\sum^{\infty}_{k=1}\frac {\sin(2\pi k\tilde\omega-\pi
/4)}{k^{3/2}},
\end{equation}
where $\tilde\omega=(\mu^2-\tilde M^2)/(2eH)$, and $\tilde
M\equiv
M(\mu)$ is the quark mass at $H=0$, $\mu\neq 0$.

\underline{\bf Conclusions:} Let us point out once more that for
strongly correlated fermionic systems there is a possibility
to observe nonperiodic magnetic oscillations.
Moreover, in such  systems in the presence of an external
magnetic field some dynamical quantities (for example,
fermion masses) should oscillate as well.  Our results 
may be applicable in astrophysics, in the physics of neutron
stars etc, where one should take into account the relativistic
character of different phenomena.

Note that the strength of the surface magnetic field of a
neutron
star is about $10^{12}$ G and in the interior it is probably
$10^{18}$ G \cite{shap}. Our numerical estimates of the
$H_{s_0}$ values using $\Lambda = 700~ \rm Mev$ show that
the magnetic field corresponding to the tricritical point $s_0$
 varies in the interval
$10^{17}$ G$\div$ $10^{18}$ G, when $1.01<G/G_c<1.2$.  Hence,
the typical neutron star magnetic field strengths are much
smaller, than the value of $H_{s_0}$, and are located in the
oscillation region of the NJL model (see Figure 1). 
So, the $H$-dependency of different physical parameters
(such as particle density, magnetization, quark mass, etc)
inside
neutron stars possibly has a nonperiodic oscillating character.

Despite the relativistic character of our investigations, we
believe that qualitatively the presented results are valid for
nonrelativistic electronic systems, and may be applicable in
condensed matter physics, too.

More complete information about phase structure as well as 
magnetic oscillations in several NJL-type models one can
find in our recent paper \cite{ekvv}.

\section*{Acknowledgments}

We would like to thank David Blaschke for useful discussions.
One of the authors (D.E.) gratefully acknowledges the kind
support and warm hospitality of the colleagues of the Theory
Division at CERN.  This work was supported in part by the
Russian Fund for Fundamental Research, project 98-02-16690, and
by DFG-project 436 RUS 113/477.

\end{document}